# Realization of a quadrupole topological insulator phase in a gyromagnetic photonic crystal

Peiheng Zhou,[1] Gui-Geng Liu,[2,#] Zihao Wang,[2] Yuan-Hang Hu,[1] Shuwei Li,[1] Qindong Xie,[1] Yunpeng Zhang,[1] Xiang Xi,[3] Zhen Gao,[3] Longjiang Deng,[1,*] and Baile Zhang,[2,4,‡]

[1]National Engineering Research Center of Electromagnetic Radiation Control Materials, State Key Laboratory of Electronic Thin Film and Integrated Devices, University of Electronic Science and Technology of China, Chengdu 611731, China

[2]Division of Physics and Applied Physics, School of Physical and Mathematical Sciences, Nanyang Technological University, 21 Nanyang Link, Singapore 637371, Singapore

[3]Department of Electronic and Electrical Engineering, Southern University of Science and Technology, Shenzhen 518055, China.

[4]Centre for Disruptive Photonic Technologies, The Photonics Institute, Nanyang Technological University, 50 Nanyang Avenue, Singapore 639798, Singapore

The field of topological photonics was initiated with the realization of a Chern insulator phase in a gyromagnetic photonic crystal (PhC) with broken time-reversal symmetry ($T$), hosting chiral edge states that are topologically protected propagating modes. Recent advances in higher-order band topology have discovered another type of topological state, as manifested by those modes localized at the corners of a sample, which are known as corner states. Here we report the realization of a quadrupole higher-order topological insulator phase in a gyromagnetic PhC, induced by a topological phase transition from the previously demonstrated Chern insulator phase. The evolution of the boundary modes from propagating chiral edge states to localized corner states has been characterized by microwave measurements. We also demonstrate topological bound states in the continuum, when the gyromagnetic PhC is magnetically tuned. These results extend the quadrupole topological insulator phase into $T$-broken systems, and integrate topologically protected propagating and localized modes in the same platform.

Topological photonics is a subfield of photonics that explores the topological states of light [1-4]. This field was initiated more than ten years ago with a gyromagnetic photonic crystal (PhC) that exhibits the Chern insulator (CI) phase with broken time-reversal symmetry ($T$) [5-7]. When biased by external magnetic fields, the gyromagnetic PhC opens a topological band gap characterized by the non-zero Chern number, hosting chiral edge states that are unidirectionally propagating modes along the one-dimensional (1D) edges of a sample [5-9]. Such photonic chiral edge states are robust against defects and disorders, being technologically promising in various applications such as topological lasers [10-12] and photonic circuits [7, 13].

Higher-order band topology is a recent breakthrough in topological physics [14-16]. In two-dimensional (2D) geometries, the higher-order topology, such as in the recently demonstrated quadrupole topological insulator (QTI) [14,17-21], will induce zero-dimensional (0D) corner states at the corners of a sample. This phenomenon has been demonstrated in various metamaterial or PhC platforms ranging from electromagnetic waves [17, 20-22] to acoustics waves [18, 23, 24] and circuits [19]. However, most previously observed corner states have been limited to $T$-invariant systems. One exception is the recent report of corner states in $T$-broken honeycomb lattice PhC with one sublattice magnetically biased [25]. However, the phase transition between chiral edge states and corner states has not been fully revealed and demonstrated.

Recent theory has predicted that the higher-order topology can potentially occur in a $T$-broken QTI system [26], unlike early QTI proposals that must require $T$ as a key ingredient. It has been proposed that gyromagnetic PhC can host the QTI phase that is induced by external magnetic fields. The benefit of realizing the QTI in a gyromagnetic PhC is two-fold. Firstly, it has the potential to realize topological phase transition between CI and QTI, integrating topologically protected propagating and localized states in the same platform. Secondly, previous QTI models in general require negative coupling that is difficult for a PhC, but a gyromagnetic PhC with broken $T$-symmetry does not rely on negative coupling.

Here, we report on the realization of the QTI phase in a gyromagnetic PhC. The topological phase transition between the QTI and the previously demonstrated CI phase is demonstrated with transmission measurements and field mappings in the microwave regime. Magnetic tuning applied to the gyromagnetic PhC also provides a unique degree of freedom for the corner states to migrate into bulk bands, forming topological bound states in the continuum. Our work has extended QTI phases into $T$-broken systems, and connected corner states with chiral edge states in the same platform.

The designed gyromagnetic PhC is depicted in Fig. 1(a). The unit cell consists of four quarters of a gyromagnetic ferrite cylinder with diameter $d$, each of which is located at a corner of a square with side $a$. A uniform magnetic field $B$ is applied along the $z$ direction to break the $T$. The entire PhC is sandwiched between two parallel metal planes, and the electric fields are polarized along the *z* direction.

As *B* and *d* vary, both the CI phase and the QTI phase can emerge in the gyromagnetic PhC, as shown in the phase diagram plotted in Fig. 1b. Here we fix *B* = 0.3 T, and plot the band dispersions of the PhC at *d* / *a* = 0.27 (C1), 0.33 (D1), and 0.38 (Q1) in Figs. 1(c)-(e), respectively. The calculated gap Chern number (*C*) and quadrupole moment ($q_{xy}$) have been indicated in the same figure (see Supplemental Material [27] for their calculation).

As shown in Fig. 1(c) for the C1 state, a complete band gap occurs between the second and the third bands, and its corresponding non-zero gap Chern number *C* = −1 manifests C1 as a CI phase. By increasing *d* to reach the D1 state (Fig. 1(d)), the gap closes and form a single Dirac point at the centre of the Brillouin zone (Γ point) [28] (see Supplemental Material [27] for the Berry curvature verification of the single Dirac point), which can be effectively treated as a magneto-optical near-zero index medium [29]. Further increasing *d* to reach the Q1 state (Fig. 1(e)), a complete gap reappears between the second and the third bands, featured by a trivial Chern number yet a nontrivial quadrupole moment $q_{xy}$ = 1/2, indicating the Q1 state as a QTI phase. During the phase transition process between the CI phase (C1) and the QTI phase (Q1), a band inversion occurs between the second and third bands, evidenced by exchanged eigenstates at the Γ point (marked as purple and red triangles in Figs. 1(c) and (e); see Supplemental Material [27] for the profiles of eigenstates). On the contrary, previously demonstrated trivial-nontrivial QTI phase transition relies on the band inversion at the M point [26]. The CI-QTI phase transition can also be induced by tuning the external magnetic field strength *B* (see Supplemental Material [27] for the magnetic field induced phase transition and band inversion).

Next, we study the boundary modes of the C1 and Q1 states by constructing two finite samples, as plotted in Figs. 2(a) and (d), respectively. Here, semi- and quarter-cylinders near the boundaries of the samples are replaced by full circular cylinders with preserved areas of cross sections to facilitate the experimental fabrication. Note that simulation results have shown that such perturbations have no essential influence on the eigenmodes and experimental observations (see Supplemental Material [27] for the materials and experimental set-ups). Perfect electric conductors (PECs) are introduced to all boundaries to prevent wave leakage into the surrounding environment. Figure 2(b) has plotted the eigenmodes solved for the C1 sample in Fig. 2(a). Clearly, the gap of the bulk states has been populated with chiral edge states, indicating the C1 sample lying in the CI phase. Putting a source antenna near a corner of the C1 sample (label "1" in Fig. 2(c)), edge states are found propagate counterclockwise along the boundaries with negligible reflection; the decay of the propagating waves results from unavoidable material losses. For the Q1 sample, Figure 2(e) shows that only four degenerate corner modes exhibit in the bulk gap, proving the emergence of a QTI phase. The corner excitation in Fig. 2(f) thus results in highly localized corner states.

Experimentally, the yttrium iron garnet (YIG) ferrite cylinders are employed to construct the C1 / Q1 sample in Fig. 3(a) / (d). The samples are covered with copper claddings to function as PECs. We measure the transmissions of the bulk, edge, and corner states by placing a pair of antennas into the fabricated gyromagnetic PhCs, as illustrated by labels "1" and "2" in Fig. 3(a). As shown in Fig. 3(b), the measured edge state transmission for the C1 sample dominates around 8.96 GHz - 9.27 GHz, in consistent with the eigensolver predictions in Fig. 2(b). Besides, we also put a source antenna near a corner of the sample and scan the excited field distribution. When the oscillating frequency lies in the bulk gap (9.12 GHz), the measured electric fields in Fig. 3(c) propagate robustly along the counterclockwise direction, in agreement with the simulation results in Fig. 2(c).

For the measured transmissions in Fig. 3(e) for the Q1 sample, both the bulk and edge states perform obvious dips around 8.04 GHz - 8.38 GHz, indicating the absence of the edge states in the bulk gap. However, a significant peak of the corner state transmission appears at 8.23 GHz, manifesting the emergence of corner states. Note that a negligible frequency shift of 0.02GHz comparing to the simulation results in Fig. 2(e) is caused by fabrication errors in experimental setups. Further field scanning results have visualized the strongly localized corner states directly, as shown in Fig. 3(f). Those results have experimentally corroborated the phase transition between a CI phase with chiral edge states and a QTI phase with localized corner states in a gyromagnetic PhC platform.

Besides, the corner states embedded in the continuum [30], a unique feature enabled by the higher-order topology, have also been realized in our platform. In previous proposals of *T*-invariant higher-order topological phases, the shift of corner states can be realized by modifying material and structural parameters [22, 24]. In our platform, the eigenfrequencies of the corner states can be simply shifted by tuning the biasing magnetic field strength *B*. Here, we fix the PhC to be *d* / *a* = 0.38 and vary *B* from 0.18 T to 0.42 T. As shown in Fig. 4(a), the band gap separates the bulk continuum into the lower and higher frequency parts. Increasing *B* continuously shifts the corner states from the higher frequency part into the gap, and then into the lower frequency part. This property has been experimentally verified by measuring the bulk and edge transmissions for *B* = 0.18T, 0.26T, 0.34T, and 0.38 T, as shown in Figs. 4(b)-(e), respectively. Specifically, for *B* = 0.38 T in Fig. 4(e), the peak of the corner transmission emerges at 8.51 GHz, merged in the lower part of the bulk states. Consequently, a source antenna at 8.51 GHz will simultaneously excite both the bulk and corner states in a finite sample, as verified by the measured and simulated field distributions in Fig. 4(f).

We have thus realized a QTI phase in a gyromagnetic PhC and revealed its phase transition with the CI phase. Both the in-gap and bulk-embedded corner states have been observed. These results expand the scope of higher-order topological physics to include *T*-broken systems. The integration of chiral edge states and corner states in the same photonic platform provides an opportunity to switch between topologically protected propagating and localized photonic modes as expected.

#guigeng001@e.ntu.edu.sg
*denglj@uestc.edu.cn
‡blzhang@ntu.edu.sg

This work is supported by the National Natural Science Foundation of China (NSFC) (No.52022018 and 52021001). Work at Nanyang Technological University is sponsored by the Singapore National Research Foundation Competitive Research Program (grant no. NRF-CRP23-2019-0007) and Singapore Ministry of Education Academic Research Fund Tier 3 (grant no. MOE2016-T3-1-006).

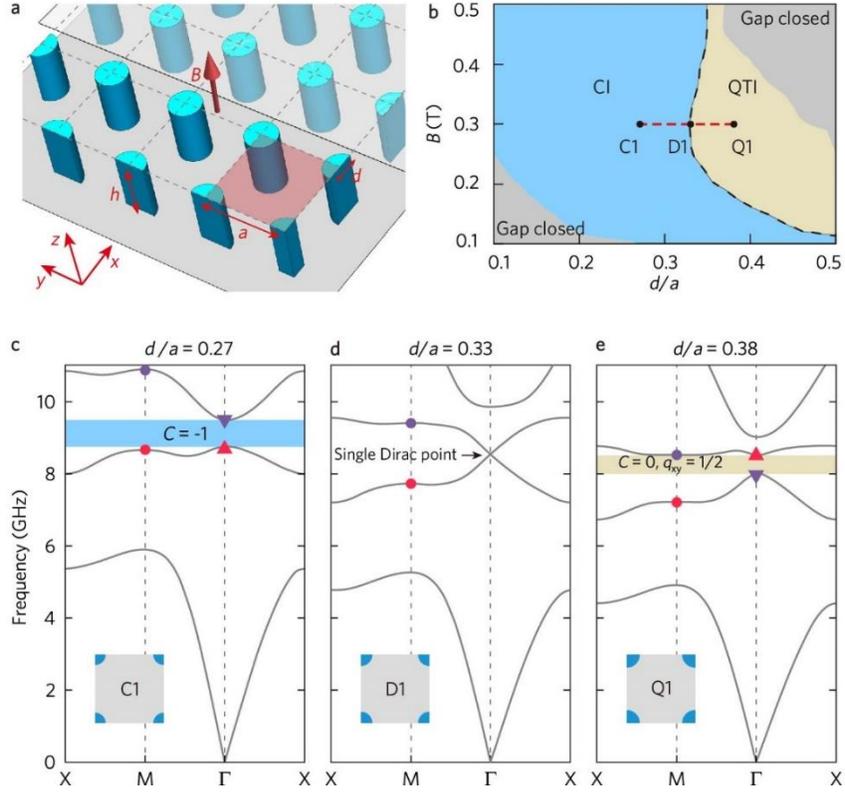

FIG. 1. Design of a gyromagnetic photonic crystal with topological phase transition from Chern insulator (CI) phase to quadrupole topological insulator (QTI) phase. (a) A photonic square lattice consisting of gyromagnetic quarter-cylinders with diameter $d$ and height $h = 8$ mm. The lattice constant is $a = 14.3$ mm. An external magnetic field $B$ is applied along $+z$ direction. The lattice is placed between two metal planes (in grey) with the upper plane half-covered and transparent for visualization. (b) Phase diagram. The topological phase transition occurs along the dashed black line. Black dots indicate three distinct phases at $B = 0.3$ T: C1 is a CI phase with gap Chern number $C = -1$; D1 denotes the phase transition point; Q1 has QTI phase of quantized electric quadrupole moment $q_{xy} = 1/2$. Photonic bandgap also closes at grey regions. (c-e) Band structure evolution induced by $d/a$ at $B = 0.3$ T. Colored dots and triangles in the band structure diagrams indicate the band inversion of second and third eigenstates at $\Gamma$ points (see Supplemental Material [27] for the eigenstates). Insets are the unit cell configurations of each crystal.

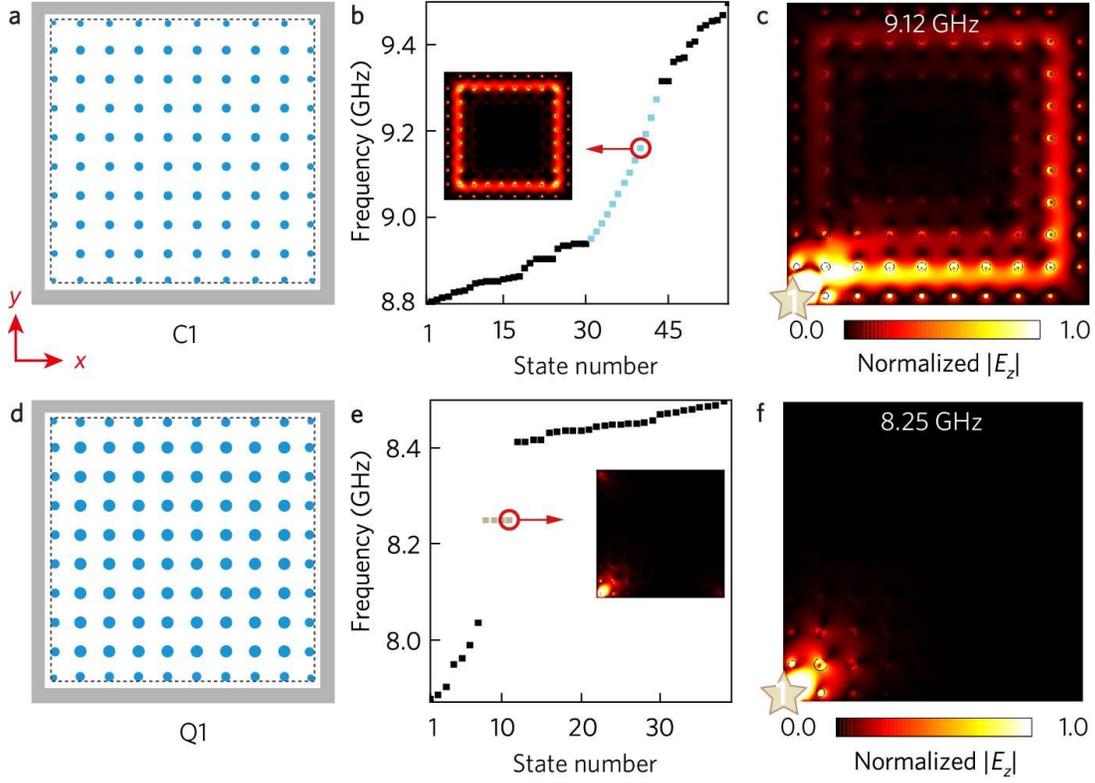

FIG. 2. Simulation of in-gap boundary states. (a) / (d) Square lattice for C1 / Q1. The samples with 9 × 9 unit cells are enclosed by PECs in simulation. Here the PEC edges are placed away from the lattice edges (black dotted lines) with a constant distance $0.42a$. (b) / (e) Numerically simulated eigenstates of C1 / Q1 sample. Edge states (blue squares in (b)) and four corner states (tan squares in (e)) are found in the bandgap spectrum with one typical profile shown in the insets. Other bulk states are colored in black. (c) / (f) Simulated $E_z$ field distributions for corner excitation of C1 / Q1 sample. Colored stars denote the source antenna (labelled as "1").

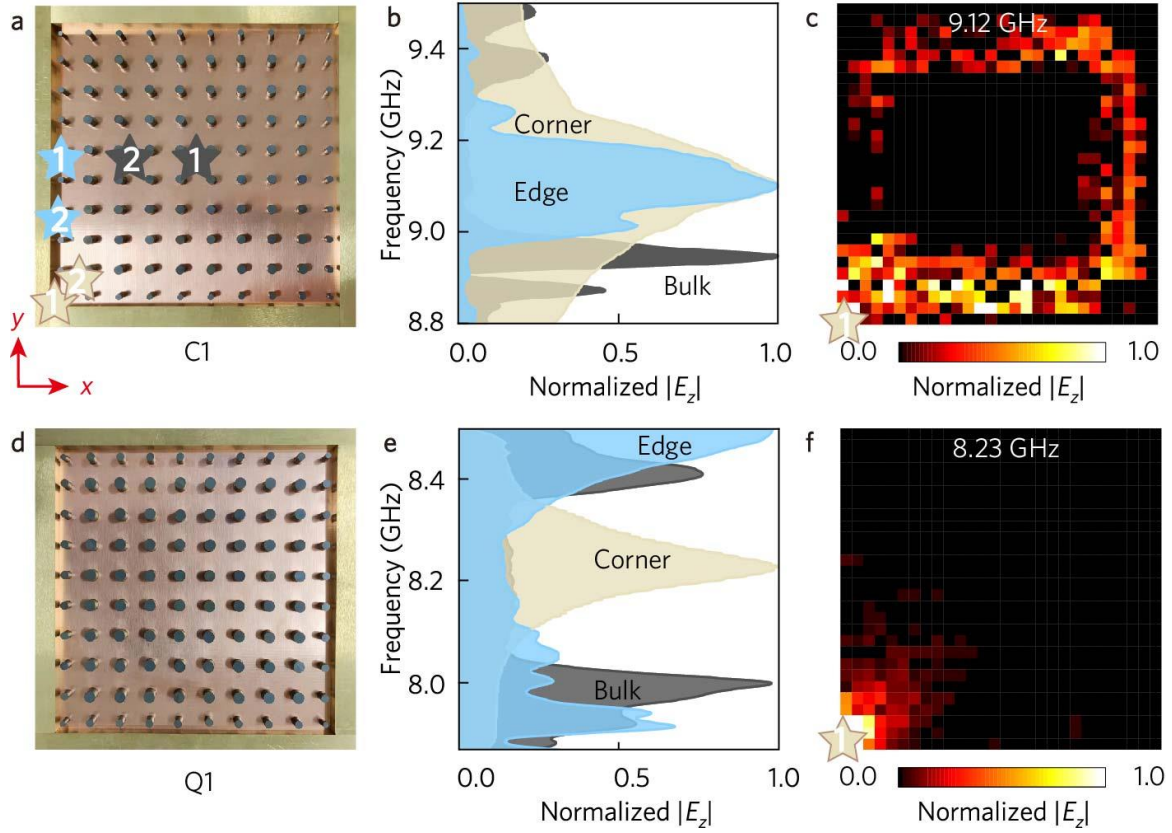

FIG. 3. Observation of in-gap boundary states. (a) / (d) Photograph of a square sample for C1 / Q1. The PhC samples are enclosed by copper bars with the same geometrical setups indicated in Fig. 2 (a) / (d). Colored stars denote the source antenna (labelled as "1") and probe antenna (labelled as "2") in three measurements for corner, edge, and bulk transmissions. (b) / (e) Measured frequency-resolved transmission for C1 / Q1 sample. (c) / (f) Measured field distributions excited by port "1" for C1 / Q1 sample.

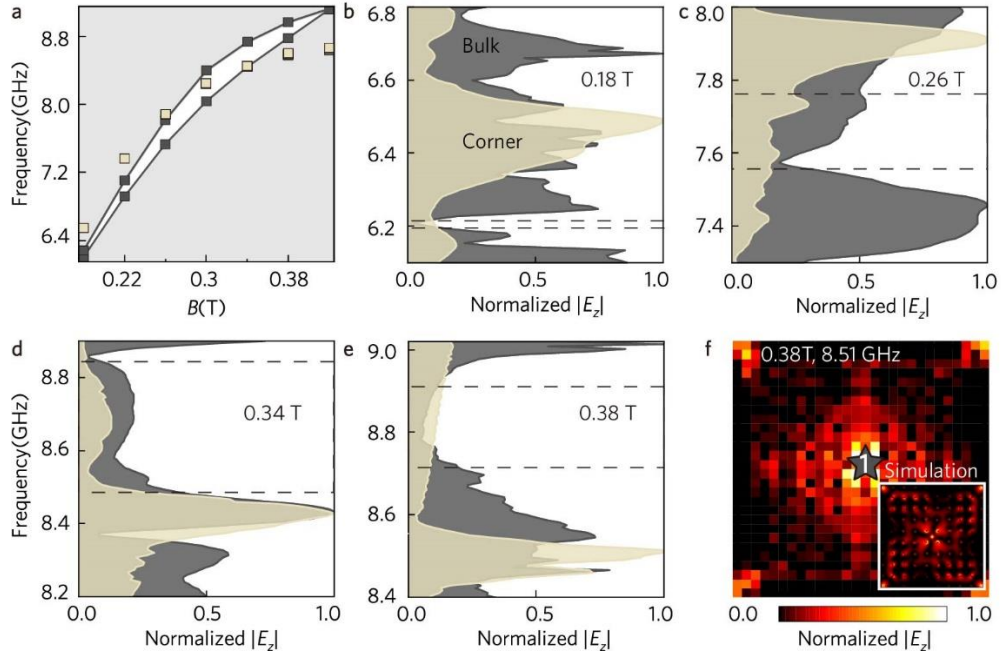

FIG. 4. Observation of corner states in the bulk continuum. (a) Numerically simulated eigenstates of QTI sample with $d/a = 0.38$. Grey / white region denotes the bulk continuum / band gap. Tan squares represent corner states. (b-e) Measured bulk and corner transmission spectra at 0.18 T, 0.26 T, 0.34 T, and 0.38 T, respectively. (f) Simulated and measured field distributions excited by port "1" at 0.38 T and 8.51 GHz.